\documentclass[aps,pre,reprint,floatfix]{revtex4-2} % Keep only this line

\usepackage{ragged2e} 
\usepackage{graphicx}
\usepackage{xcolor} 
\usepackage{amsmath, amssymb}
\usepackage[colorlinks=true,linkcolor=blue,urlcolor=blue,citecolor=blue]{hyperref}
\usepackage{times}
\usepackage{bm}
\usepackage{nameref}
\usepackage{array}
\usepackage{colortbl}
\usepackage{caption} 
\usepackage{float}
\usepackage{siunitx}
\usepackage{xcolor} 
\usepackage{comment}
\usepackage{braket}
\usepackage{amsfonts}
\usepackage{hyperref}
\usepackage{booktabs}
\usepackage{caption}
\usepackage{lipsum}
\usepackage[utf8]{inputenc}
\usepackage[T1]{fontenc}
\usepackage{amsmath} 
\usepackage{colortbl}
\usepackage{enumitem}
\usepackage{xcolor}
\definecolor{RED}{rgb}{1,0,0} % defines RED as pure red
\usepackage{tikz}
\usepackage{soul} 
\usepackage{ulem}
\providecommand{\tightlist}{%
  \setlength{\itemsep}{0pt}\setlength{\parskip}{0pt}}
\begin{document}

\title{Quantum Computing Inspired Approach for Self-Avoiding Walk (SAWs):  2D lattice and 3D lattice SAWs for single chain enumeration}

%\author{Hemant Mishra$^{1\dagger}$, Shubham Singh$^{2}$, Rajeev Singh$^{2}$, Amit Raj Singh$^{1*}$}

%\affiliation{\setcounter{affil}{0}%
%$^{1}$Department of Physics, Graphic Era Hill University, Dehradun, 248002, India\\
%%$^{2}$Department of Physics, Indian Institute of Technology, Banaras Hindu University, Varanasi,221005,India}
%\email{amitsakr@gmail.com\\
%$\dagger$ hemant.ev@gmail.com}
\author{Hemant Mishra$^{\dagger}$}
%\thanks{Present address: Department of Physics, Graphic Era Hill University, Dehradun 248002, India}
\affiliation{Department of Physics, Graphic Era Hill University, Dehradun 248002, India}

\author{Shubham Singh}
\affiliation{Department of Physics, Indian Institute of Technology (BHU), Varanasi 221005, India}

\author{Rajeev Singh}
\affiliation{Department of Physics, Indian Institute of Technology (BHU), Varanasi 221005, India}

\author{Amit Raj Singh}
\email{amitsakr@gmail.com\\
$\dagger$ hemant.ev@gmail.com}
\affiliation{Department of Physics, Graphic Era Hill University, Dehradun 248002, India}
\begin{abstract}
We investigate the application of quantum computing algorithms to enhance the efficiency of enumerating self-avoiding walks (SAWs), utilizing quantum properties such as superposition and interference. A Quantum Amplitude Estimation (QAE)-based algorithm is developed to enumerate SAWs on both 2D and 3D lattices. In case of 2D square lattice, SAWs up to N=71 steps are successfully enumerated within 26.9 minutes - significantly improving upon the classical algorithm, which required approximately 231 hours (\textcolor{blue} {Jensen et al., 2012, J. Phys. A: Math. Theor. 45, 115202}). The algorithm is further extended to 3D cubic lattices, where SAWs up to N=40 steps are enumerated in 13.06 minutes, compared to the classical result of N=36 in 250 hours (\textcolor{blue}{Schram et al., 2011, J. Stat. Mech. P06019}). These results demonstrate a substantial reduction in computational time, highlighting the potential of quantum computing for combinatorial enumeration problems.
\end{abstract}

\maketitle

\section{Introduction}
\justifying
Self-avoiding walks (SAWs), a foundation of lattice-based statistical physics research, were introduced by Paul Flory (1949). There are several applications of SAW in statistical physics research, like modelling the behaviour of polymer chains in a solvent, where the polymer’s conformations must avoid self-intersections, and the study of critical phenomena, like phase transitions, as the scaling behaviour of SAWs provides insight into the critical exponents that describe these transitions \cite{slade1994}. Moreover, SAWs are also used in the study of percolation theory, where they help to explain how connectivity and clustering evolve in disordered systems \cite{Madras1993}. The self-avoiding constraint mimics the excluded volume effect; SAW is a path on a lattice where each step moves from one lattice point to an adjacent one, and the path does not visit the same point more than once \cite{Madras1993}. Formally, an $n$-step SAW on a 2D and 3D lattice is a sequence of distinct vertices, where each pair of consecutive vertices is connected by an edge \cite{slade1994}. These constraints are also useful for studying spreading patterns in nature. \par
The asymptotic behavior of SAWs provides insights into polymer conformation, such as size (end-to-end distance), shape (radius of gyration), nearest-neighbor interactions (covalent bonds), and degeneracy \cite{Madras1993}. These insights are used to describe the folding-unfolding of protein and the denaturation of single-stranded DNA (ssDNA) or double-stranded DNA, ensuring that strands avoid self-intersection \cite{Singh2010, Singh2009, ARSingh2009}. To model the polymer on a 2D and 3D lattice, the enumeration of self-avoiding walks involves counting the number of distinct paths that can be formed under the self-avoidance constraint. In the 2D and 3D cases, the enumeration of self-avoiding walks has been a fascinating topic of considerable research interest since the 1940s \cite{Zbarsky2019}. This enumeration is frequently a challenging issue, as the number of such walks grows exponentially with the length of the walk, and finding an exact count for large numbers of steps remains challenging \cite{Kesten1963}. Although the enumeration of SAWs on 2D and 3D lattices is complex, self-avoiding walks have been studied extensively, and various techniques have been developed to approximate or find exact counts, especially for small lattice sizes \cite{slade1994}. There are several classical algorithms that have been proposed to enumerate the exact number of SAWs, including exact enumeration, and parallel computation methods. The exact enumeration of SAWs in 2D and 3D has been tackled with exact methods up to modest walk lengths, but for larger walks, Monte Carlo simulations and other approximate techniques are commonly employed \cite{Madras1993}. These approaches have limitations of their own, including a lack of precision in the number of enumerations for Monte Carlo and high time complexity for exact enumeration and parallel computation methods.
\par\vspace{5pt} % Adjust the spacing as needed
In the past several decades, there has been a lot of literature on counting SAWs. Orr et al. calculated enumeration $(Z_N)$ manually for every $N$ up to $N_{max}$ = 6 \cite{Orr1947}. Fisher and Sykes \cite{Fisher1959, Madras1993} used a computer to list every SAW in 3D up to $N_{max}$ = 9. Guttmann et al. \cite{Guttmann1987, Guttmann1989} listed longer SAWs up to $N_{max}$ = 20 , and MacDonald et al. \cite{MacDonald1992, MacDonald2000} achieved $N_{max}$ = 23 , and further, they achieved $N_{max}$ = 26. In recent years, the total number of enumerations in 3D and 2D has been calculated for maximum lengths of 36 and 79, respectively. The total count of SAWs for N=79 on a 2D square lattice was done using a new transfer matrix method, taking about 16,500 CPU hours with 400 processors, which is roughly 41.25 hours \cite{jensen2013}. The length-doubling method is a novel computational technique that accelerates the exact enumeration process compared to the traditional method. After the implementation of the length-doubling method in the computational technique, the enumeration of SAWs was calculated up to 36 steps. The enumeration up to N = 36 required 50,000 CPU hours on the Huygens supercomputer at SARA in Amsterdam \cite{Schram2011}.
\par
Enumerating the walks, especially in higher dimensions or for large lattice sizes, becomes computationally intractable for classical algorithms. The reasons are: 
\begin{enumerate}[noitemsep, topsep=0pt, after=\vspace{0.5\baselineskip}]
\item \textbf{Exponential Growth:} The number of possible self-avoiding walks grows exponentially with the size of the lattice and the length of the walk.
\item \textbf{Time Complexity:} Classical methods requires large computation time to enumerate all possible walks, this grows rapidly with the lattice size.
\item \textbf{High-Dimensional Lattices:} In higher-dimensional lattices, the number of self-avoiding walks becomes extraordinarily large, making computation more complex.
\item \textbf{Lack of Efficient Enumeration Methods:} Classical algorithms often rely on recursive or dynamic programming approaches, but these methods still become overwhelmed with large inputs, leading to high computational overhead and memory demands.
\end{enumerate}

Unlike classical computing, quantum computing offers a promising approach to solving difficult combinatorial problems, such as enumerating selfavoiding walks (SAWs) on 2D and 3D lattices. Quantum computing algorithms have the potential to significantly accelerate the enumeration process through techniques like quantum parallelism.  This can explore multiple configurations simultaneously by using quantum superposition. The fundamental unit of information in quantum computing is quantum bits, or qubits, which can exist in state 0, state 1, and simultaneously in both states \cite{sutor2019}. This special characteristic enables quantum systems to handle a large number of possibilities at once, as a result, quantum computing offers strengthening computational speed and efficiency over classical computing by enabling multiple operations to be executed simultaneously. Here, we have developed a quantum computing inspired algorithm to enumerate the SAWs on 2D and 3D lattices, which achieves significantly faster performance than classical algorithms for large step count.
 
\section{Quantum Computing Inspired Algorithm Implementation}
\label{sec:QAE Impl}
Quantum computing works on the principle of superposition and interference. These principles of quantum computing are used to design an efficient quantum algorithm, which shows a demonstration of a reduction in time complexity to find a marked item among the M items.\cite{sutor2019} \\
\par
In this paper, we present the well-known quantum algorithm \textbf{Quantum Amplitude Estimation (QAE)}  specifically designed for the enumeration of large-step self-avoiding walks (SAWs).  In the context of reducing time complexity in comparison to classical algorithms and achieve quadratic speedup, the QAE algorithm is designed for a bipartite system \cite{grinko2021}, which is encoded using Grover's algorithm \cite{Stoudenmire2023}.
By using a quantum computing framework, we decompose the algorithm into several interrelated components, each addressing a specific aspect of the overall computation. The design begins with the state preparation for uniform superposition over all states using a quantum encoding scheme that captures the lattice structure and enforces self-avoidance through a quantum Oracle to mark the valid SAWs. Next, we define a quantum walk operator tailored to evolve the system across valid paths, embedding constraints directly into the unitary operations. To maintain the feasibility of the quantum walk operator over longer walks, we introduce a filtering mechanism based on the suppression of amplitudes of invalid walks and the amplification of valid walks. This pruning mechanism is known as the quantum amplitude estimation. It is encoded with the Grover search that leverages inverse QFT to unravel the phase of states \cite{Mandviwalla2018}. The efficiency is preserve through QAE via ancilla qubits. For longer walk qubit experiences decoherence which is mitigated by a technique of zero-noise extrapolation \cite{Kandala2019}. \\
A hybrid simulation strategy was also employed to address the limitation for large scale qubits simulation on classical hardware. First, IBM's statevector simulator provide an baseline framework to evolve the entire quantum register into smaller tractable subregister. At the relevant point in the computation, qubits that had accomplished their contribution in SAWs enumeration were reinitialized and compressed into tensor representation. This approach efficiently manage the entanglement growth and qubit correlations in tensor-network style \cite{tensor2023} which preserves the correctness of SAWs enumeration. These techniques are effective to bypass the growing obstruction.\\

\begin{figure}[!htbp]
    \centering
    \includegraphics[width=0.9\linewidth]{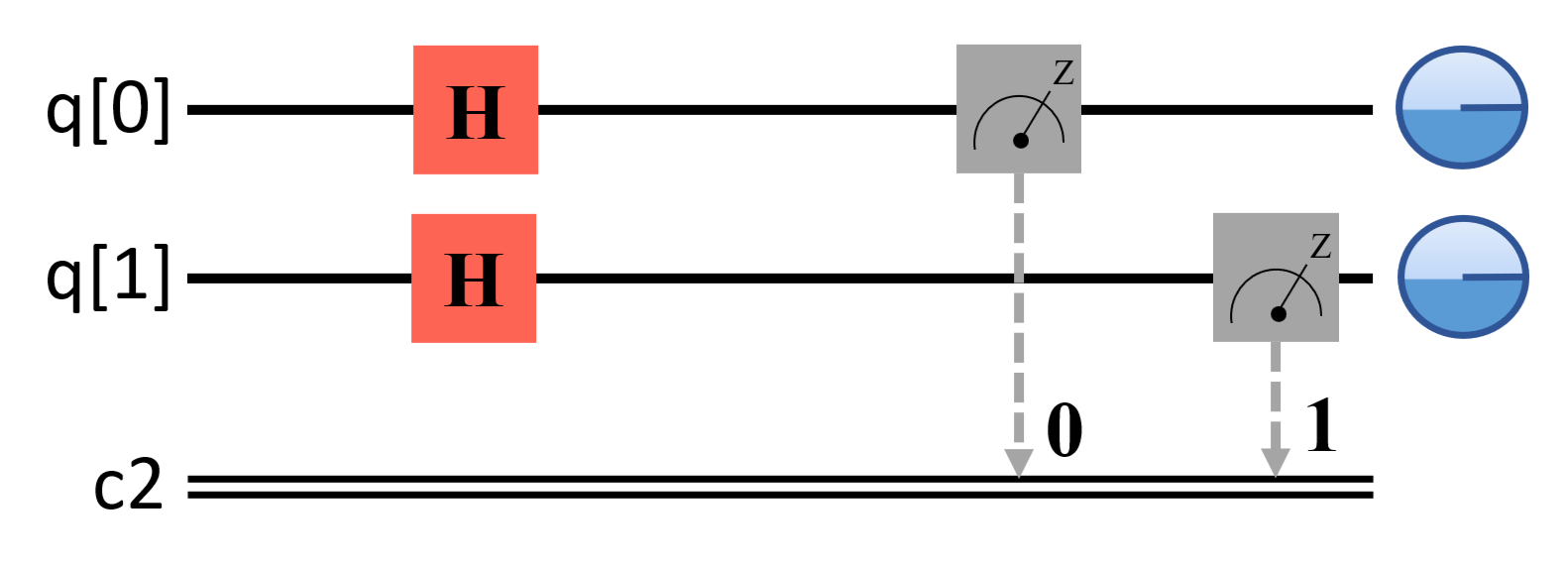} % Adjust width as needed
    \caption{Quantum circuit for state preparation of $N=1$ step SAW, with 2 qubits over $4^1$ = 4 configurations. Red boxes $(H)$ represent Hadamard Gates for initializing superposition. Grey boxes represent the Z-basis for the measurement at the end of the computation.}
    \label{fig:1}
\end{figure}

\textbf{State Preparation:} 
We begin with uniform superposition over all probable walks, where each step requires $2$ qubits, with input length N in 2D square lattice and in 3D square lattice each step requires $3$ qubits to validate the available direction corresponding to steps. Figure~\ref{fig:1} demonstrate the quantum superposition through a quantum circuit for 1 step SAWs in 2D. (for details see Appendix \ref{app:superposition})

For 2D SAWs on a square lattice, each step can move in available four directions, i.e. right (+X), left (-X), up (+Y), down (-Y) all these possibilities encoded with two qubits corresponding to their basis states i.e., $00$: +X(move right), $01$: -X(move left), $10$: +Y(move up), $11$: -Y(move down). 
Thus, for \(N\)-step walk it requires 2N qubits for full encoding. The superpostion of quantum states can be represented as;
\begin{equation}
|\psi\rangle = \sum_{i=0}^{2^{2N}-1} \alpha_i |i\rangle
\end{equation}
where, $|i\rangle$ represents the set of sequences corresponding to the basis state for specific walk configuration in 2D  and $\alpha_i$ is the amplitudes of basis states $(\alpha_i = \frac{1}{\sqrt{4^N}})$. (see Appendix \ref{app:amplitude}) \\

\begin{figure}[!htbp]
    \centering
    \includegraphics[width=0.8\linewidth]{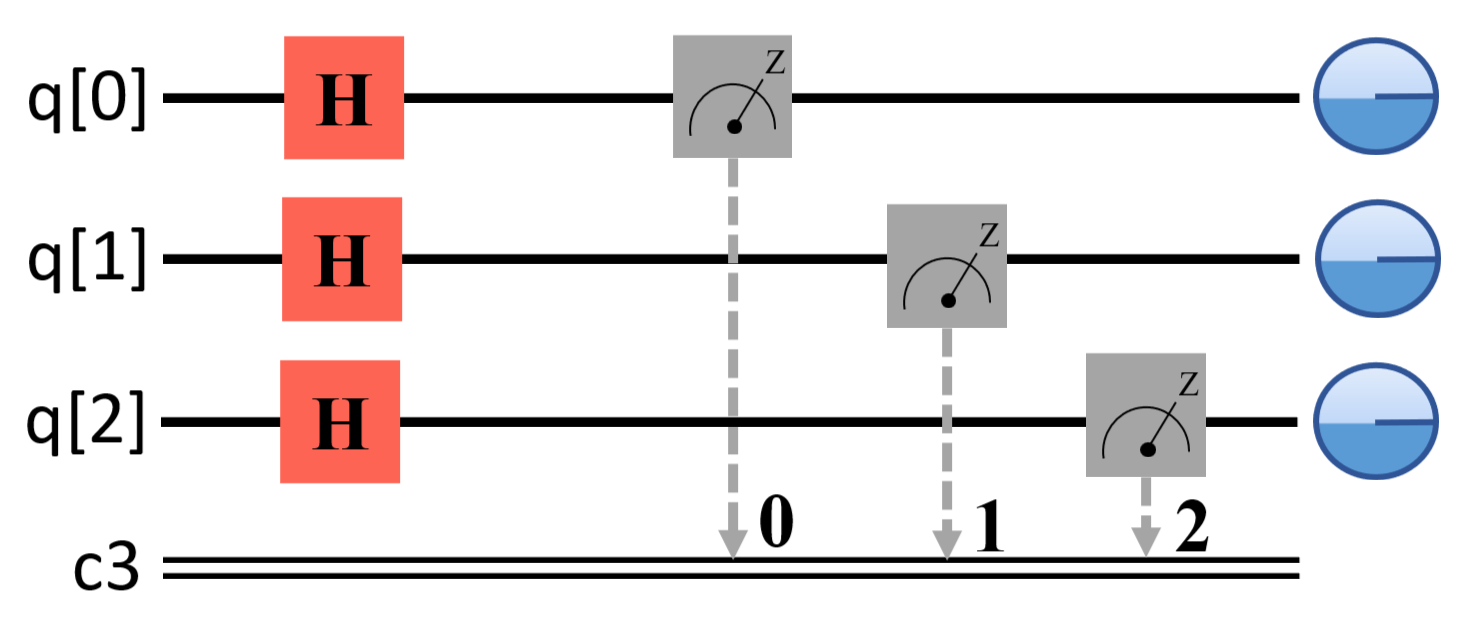} % Adjust width as needed
    \caption{Quantum circuit for state preparation of $N=1$ step SAW, with 3 qubits over $6^1$ = 6 configurations. Red boxes $(H)$ represent Hadamard Gates for initializing superposition. Grey boxes represent the Z-basis for the measurement at the end of the computation.}
    \label{fig:2}
\end{figure}

Similarly, for 3D SAWs on a simple cubic lattice, Figure~\ref{fig:2} demonstrates the quantum superposition through a quantum circuit for 1-step SAWs. Here, the algorithm follows the same fundamental principles but encoding requires additional qubits (total 3 qubits instead of 2 qubits as in the 2D lattice) to represent the six available directions right (+X), left (-X), up (+Y), down (-Y), forward (+Z), backward (-Z). The basis states  in 3D lattice  are $|000\rangle$: +X(move right), $|001\rangle$: -X(move left), $|010\rangle$: +Y(move up), $|011\rangle$: -Y(move down), $|100\rangle$: +Z(move forward), $|101\rangle$: -Z(move backward). 
Thus, for \(N\)-step walk, it requires 3N qubits for full encoding. The  superposition of quantum states can be represented as
\begin{equation}
|\psi\rangle = \sum_{i=0}^{2^{3N}-1} \alpha_i |i\rangle
\end{equation}
where, $|i\rangle$ represents the set of sequences corresponding to the basis state for specific walk configuration and $\alpha_i$ is  the amplitudes of basis states, $(\alpha_i = \frac{1}{\sqrt{2^{3N}}})$.(see Appendix \ref{app:QAE})\\

\textbf{Quantum Oracle Implementation:}
Quantum oracle is a critical component of the algorithm, responsible to marking the states corresponding to valid SAWs, and verifying SAWs constraints \cite{Oracle1994} (i.e., N steps requires $O(\sqrt{M})$ operation, where $M$ represents number of oracle calls). In the quantum oracle implementation, the amplitude $\alpha_i$ is normalized to ensure the total probability of all possible outcome in a quantum system sums to $1$ to verify pure self-avoidance (shown in the Appendix \ref{app:amplitude})
\[
\alpha_i = 
\begin{cases} 
\frac{1}{\sqrt{|V|}} & \text{if } i \in V \text{ (valid direction sequence)} \\
0 & \text{if } i \notin V \text{ (invalid direction encoding)}
\end{cases}
\]
where, $i$ represents the set of sequences corresponding to the basis state and $V$ represents set of indices for valid direction sequences (and $|V|$ represents its cardinality). This ensures $V$ to project onto the valid subspace i.e. $\sum_i|\alpha_i|^2=1$. Quantum oracle perform consistently for small N-step walk, however, for large N-step walk it becomes computationally intensive due to overhead marking of valid and invalid SAWs, as well as parsing memory error and processing constraints. To address this issue, the Quantum Amplitude Estimation algorithm is implemented.\\

\textbf{QAE Mechanism:}
Quantum Amplitude Estimation offers quadratic speedup. It applies Phase Amplification for valid SAWs across all possible walk configurations (which consist of valid and invalid SAWs). This QAE algorithm is integrated with Inverse Quantum Fourier Transform $\text{QFT}^\dagger$ \cite{pittenger2012} that encodes the Grover's search to validate the marked oracle states for the counting of valid SAWs. For estimating the probability of generating valid SAWs we established the equation~\eqref{eq:QAE} (see Appendix \ref{app:QAE})
\begin{equation}
    A|0\rangle = \sqrt{a}|1\rangle|\psi_1\rangle + \sqrt{1-a}|0\rangle|\psi_0\rangle 
\label{eq:QAE}
\end{equation}
where,
$\textbf{A:}$ A quantum operator that produces the complete superposition and execute for the resulting state, \textbf{$|1\rangle|\psi_1\rangle$}: Produces ``true'' state corresponding to valid SAWs, \textbf{$|0\rangle|\psi_0\rangle$}: Produces ``false'' state corresponding to invalid SAWs, and $\textbf{a}:$ represents the probability of generating valid SAW (i.e., $a = Z_N / 4^N $) where, $Z_N$ number of SAWs of length N and $4^N$ shows total number of possible walks. This value of $a$ is quantified through the grouping of flag qubits. (mentioned in Appendix \ref{app:QAE}).\\
From the Grover Search algorithm, we only use its Grover's iteration to leverage $QFT^\dagger$ for sustained efficiency of QAE. The primary objective of this iteration is to enhance the likelihood of achieving the desired outcome. This iteration is represented as;
\begin{equation}
   \text{iterations} = \frac{\pi}{4\arcsin(\sqrt{a})} 
\end{equation}
where ${a}$ is the probability amplitude of the desired outcome, the ${arcsine}$ function is integrated because Grover's algorithm rotates the quantum state towards the target state. Whereas the QAE iteration \cite{grinko2021} requires for oracle call $\mathcal{O} \left( \frac{M}{\epsilon} \right)$ where $M$ is the search space size and $\epsilon$ represents the precision for the desired outcome. Moreover, QAE is based on the phase estimation framework. The $QFT^{\dagger}$ has been used to extract the amplitude $a$ and also used to convert the phase information of the quantum register into a bitstring.
To store the phase information in a quantum register, an auxiliary qubit is encoded. The quantum register for ancilla qubits encodes the phase $\theta$ to resemble a Fourier series i.e., $\theta=arcsin(\frac{\sqrt{a}}{\pi})$. The circuit resemblance for $QFT^\dagger$ is obtained through the following circuits, which are constructed by using, Hadamard Gate that are applied to each qubit for overall superposition, Controlled Phase Rotation Gate for phase shift whose matrix representation is;
\[
R_k = \begin{pmatrix} 1 & 0 \\ 0 & e^{-2\pi i / 2^k} \end{pmatrix}
\]
and Swap Gate to reverse the qubit order. 

\begin{figure}[!htbp]
    \centering
    \includegraphics[width=1.0\linewidth]{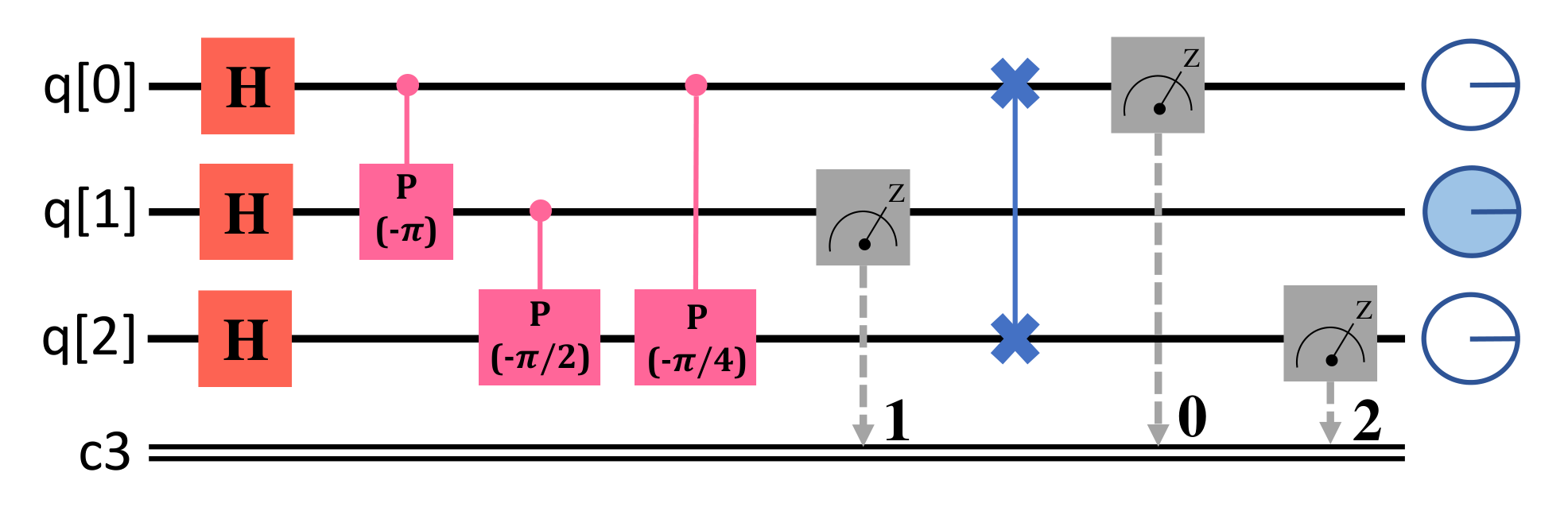} % Adjust width as needed
    \caption{Quantum circuit for the fundamental function of $QFT^{\dagger}$ through Hadamard Gate $(H)$, Controlled Phase Shift Gate , and Swap Gate}
    \label{fig:3}
\end{figure}

In matrix $R_k$ the element $e^{-2\pi i / 2^k}$ is the eigenvalue of the Q-operator in Grover's Search i.e.,
\[
Q = -A S_0 A^{-1} S_{\text{true}}
\]
where $A$ is the State preparation circuit and $S_0$ = $I-2|0\rangle\langle0|$ this shows reflection about initial state and $S_{true}$ = $I-2|1\rangle\langle 1|$ shows reflection over true state.\\

\textbf{Time Complexity}
The core work of the QAE algorithm is to reduce the time complexity with efficient computation. For classical 2D SAWs, \cite{jensen2013}, the time complexity is $\mathcal{O}(W)$, where $W$ represents total number of distinct self-avoiding walks of length N.\\
Our QAE algorithm offers reduced time complexity due to quadratic speedup, which for quantum-based 2D SAWs is
\begin{equation}
    \mathcal{O}\left(\sqrt{\mu^{N}}\right)
    \label{eq:2dTC}
\end{equation}
where, $\mu$ represents the connective constant (see Appendix D for details). This equation is determined through Quantum Amplitude Estimation (QAE) with an iteration complexity of $\mathcal{O} \left( \frac{M}{\epsilon} \right)$, where $M$ represents number of oracle calls and $\epsilon$ is the maximum allowable deviation between valid and invalid values of $Z_N$. For walks with steps $(N \leq 71)$, this technique considerably reduces the time complexity.\\
However, the limitation arises for walks with steps $(N\geq 71)$ beacuse the oracle experiences a lookup overhead that varies between a constant factor of $1$ and approximately $1.643$. This oracle lookup deflection shows the complexity of checking walk validity with N, and this causes number of oracle calls to increase polynomially, which is not an ideal condition for enumerating larger walks. The choice of $N=71$ marks oracle deflection point at which its encoding exceed practical resource limits. Thus, the quantum circuit requires encoding with $2\times(N-base_{step}) + 1$ qubits, where the parameter $base_{step}$ mitigate this overhead by applying rotation shift within the quantum circuit. Resulting, the time complexity equation modified accordingly for large N.

\begin{equation}
    \mathcal{O}\left(\sqrt{\mu^{N-base_{step}}}\right)
    \label{eq:3DTC}
\end{equation}
Similarly, in a 3D lattice, the time complexity will be the same as Equation~\eqref{eq:2dTC} where $\mu$ represents the connectivity constant of the lattice. The process of encoding the time complexity for large $N$ in quantum circuits takes place in the same way as in 2D SAWs, but with an additional qubit, i.e., $3\times(N-base_{step}) + 1$ qubits. The extended enumeration is validated by the same Equation~\eqref{eq:3DTC}.

The reduction of the time complexity through iteration in a quantum circuit, transpiled via shots, which refers to the number of times a quantum circuit is executed, allowing statistics about the quantum state. The program uses $2^{20}$ shots, and QAE's advantage lies in reducing the number of oracle calls and time complexity.\\
To counter memory parsing error and processing constraints, we implement the Zero-Noise Extrapolation (ZNE) in QAE algorithm. It anticipates the measurement outcome of a noisy quantum circuit at different noise strengths to determine valid SAW. ZNE runs at evolution time $(t)$, in contrast to SAWs $``evolution-time (t)"$ govern the timeframe length in which the quantum state evolves under the Hamiltonian \cite{Kandala2019}. The longer the timeframe, the greater increase of quantum noise in the circuit due to more gate operations at one oracle call. The circuit operates at 3-evolution times (t), where $t=0.5$ \& $t=1.0$ \& $t=1.5$; these three scales refer to Less Noise, Baseline Noise, and Amplified Noise, respectively. And, this also explains how well the value corresponding to valid SAWs is extrapolated, in which it fits Quantum states with a polynomial model corresponding to the values of $Z_N$ versus $t$. It is shown with $[Z_N(t) = a.(t)+b]$ where $a$ represents coefficient of rate of change of initial $Z_N$ with respect to $t$ and $b$ work as an intercept which refers to noise-free extrapolation of the desired value at $t=0$.
\begin{figure}[!htbp]
    \centering
    \includegraphics[height=0.8\columnwidth,keepaspectratio]{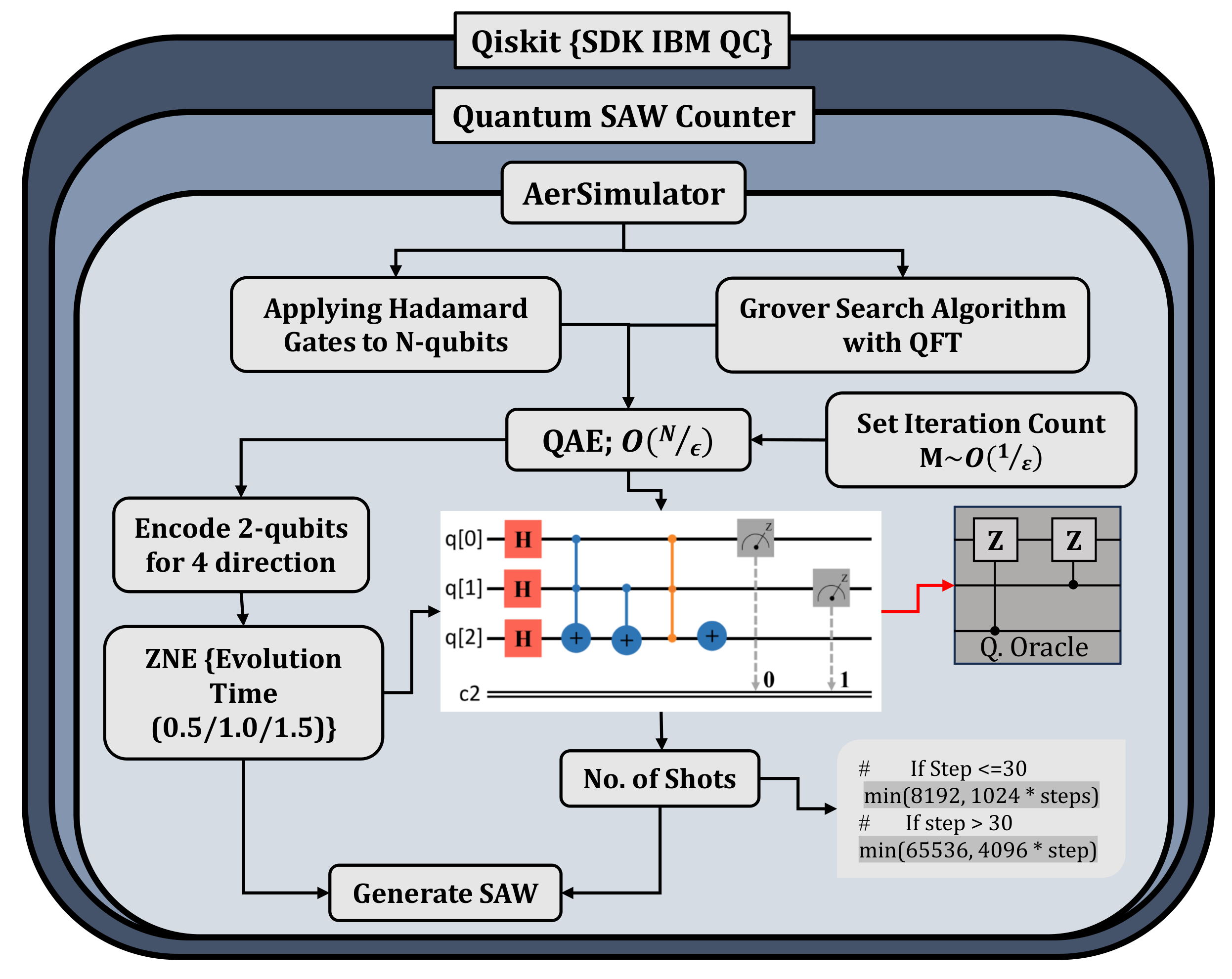} % Adjust width as needed
    \caption{Schematic Flowchart for QAE Implementation by applying superposition state with Grover's Iteration encoded with $QFT^\dagger$ that is further applied in QAE for the enumeration of N-steps.}
    \label{fig:4}
\end{figure}

To perform QAE for the enumeration of SAWs, \textbf{QISKIT} SDK of IBM Quantum Computing has been used. QAE uses AerSimulator to simulate the algorithm, which in the backend generates bitstrings representing valid walks with thread locking for safe execution \cite{Sutorpython2019}. Figure~[\ref{fig:4}] represents a schematic flowchart of the algorithm that focuses on constraining the exponential growth of time complexity. And, the number of iterations in QAE was set to balance the amplification efficiency with valid computation.
\par
The implementation of the algorithm was conducted using Python v3.12.3 with Qiskit (version 2.0.0) for quantum circuit design and simulation, Qiskit Aer (version 0.16.0), and QuTiP for rendering the representation of the Bloch Sphere executed on a Windows 11 system with Anaconda. The simulation was performed on a system with 8GB RAM and an Intel i5 multi-core processor.

\section{Result}
\subsection{2D SAW Computational Analysis}
\begin{figure}[!htbp]
    \centering
    \includegraphics[width=1.0\linewidth]{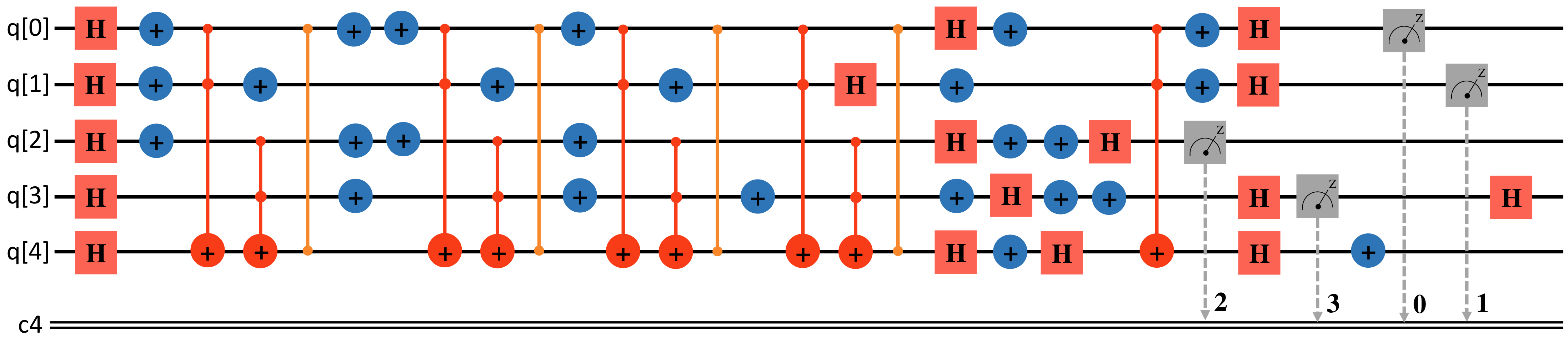} % Adjust width as needed
    \caption{Quantum Circuit demonstration for N=2 in 2D Lattice. Red boxes $(H)$ represent Hadamard Gates for initializing superposition. Blues $(+)$ indicates NOT $X$ gate for phase flip. Red $(+)$ symbols denote the Toffoli $(CCX)$ gate for controlled condition. The yellow line represents the Controlled-Z $(CZ)$ gate, which introduces a phase shift. Grey boxes represent the Z-basis for the measurement at the end of the computation.}
    \label{fig:5}
\end{figure}

From the outlined methodology, the algorithm provides an efficient result for the enumeration of SAWs. In Figure~[\ref{fig:1}], a quantum circuit is demonstrated for the enumeration of Step 1 SAWs, which is an ideal circuit. But as mentioned above, for a larger step, a quantum circuit is encoded with additional qubits, $i.e., (2N+1)$. Figure~[\ref{fig:5}] demonstrates the complete quantum circuit for $N=2$ steps to enumerate all valid walks.
Initially, 5-qubits are encoded in which the qubits from $q[0]$ to $q[3]$ encode two successive steps of the walk, where each step is represented by two bits corresponding to spatial direction  X/Y and orientation +/-, and one ancilla qubit $q[4]$ is employed  to mark valid SAWs. Initially, Hadamard Gate $(H)$ is applied at each qubit from $q[0]$ to $q[4]$ for complete superposition over all possible walks which quantify the total configuration $(4^2$ = $16)$ with valid and invalid walks both. To identify valid SAWs, a quantum oracle is implemented for each direction to detect backtracking events, such as a step in the +X direction followed immediately by a step in the  -X direction, which constitutes an invalid walk. 
The oracle is obtained using a sequence of controlled operations, including Toffoli and controlled- Z gates, which flip the phase of invalid paths while leaving valid paths unchanged. 
This phase marking enforces the SAW constraints within the quantum state space.
%To mark the valid walks, quantum oracle is implemented at each direction to verify the backtracking (e.g., moving +X then -X $\rightarrow$ invalid). This process is continued through controlled gates: Toffoli gate and Controlled-Z gate to flip the phase of invalid paths and reinforce the SAW constraints. 
%The QAE algorithm recognizes the valid walks by enhancing their amplitudes and reducing them for the invalid ones. And it also \textcolor{red}{implements} a precision factor to validate minimum deviation ($\approx{10^{-3}}$) in between states. This process is executed through Grover's iteration and $QFT^\dagger$ that works as a diffusion operator, which is responsible for the suppression of amplitude for invalid walks. Inside the quantum circuit, this process is encoded by applying the Hadamard gate $(H)$ again to bring back qubits into superposition, and the Not gate $(X)$ inverts the state of qubits, which means reflection about the mean. The execution of $N=2$ circuit gives the enumeration $Z_N=12$.\\
The Quantum Amplitude Estimation (QAE) algorithm subsequently distinguishes valid walks by amplifying their probability amplitudes while suppressing those of invalid configurations. A precision parameter is also incorporated to ensure a minimum deviation threshold of ($\approx{10^{-3}}$)  between quantum states. The amplification process is executed through Grover iterations, with the Inverse Quantum Fourier Transform ($QFT^\dagger$) acting as the diffusion operator responsible for redistributing amplitudes. Within the quantum circuit, this diffusion operation is implemented by reapplying Hadamard gates to all qubits, followed by Pauli-X gates that invert the qubit states, effectively performing a reflection about the mean amplitude. The execution of $N=2$ circuit gives the enumeration $Z_N=12$.\\
%Similarly, the same QAE-based quantum circuit framework is extended to larger systems.  For $N=71$ steps,  Hadamard Gates $(H)$ which are applied at each qubit from $q[0]$ to $q[142]$ for superpositions over all qubits, which quantify both valid and invalid walks. Quantum oracle applied as diffusion operator through Pauli-X gates $X$ for each pair of consecutive steps (i.e., step $i$ and step $i+1$) in each direction. To control the iteration process, continued through controlled gates: \textcolor{red}{we apply} 4-Toffoli gate at each steps to nullify the invalid ones. And Controlled-Z gate to flip the phase of those invalid paths. Again Hadamard gate $(H)$ and Not gate $(X)$ are applied simultaneously with a 2-Toffoli gate for Grover's iteration and $QFT^\dagger$ to amplify the amplitude of valid SAWs. These implementation computes the enumeration for $N=71$ in $26.92$ minutes shown in the Table [\ref{tab:saw_data}].\\ 
The same QAE-based quantum circuit framework is extended to larger systems. For  N=71 steps, Hadamard gates are applied to qubits q[0] through 
q[142], generating a superposition over all valid and invalid walk configurations. The quantum oracle, implemented via Pauli-X gates, operates on each pair of consecutive steps to detect 
directional backtracking. To eliminate invalid paths, four-controlled Toffoli gates are applied at each step, followed by controlled-Z gates to flip the phases of the corresponding states. Grover’s iteration is again realized by the combined application of Hadamard and Pauli-X gates, together with two-controlled Toffoli gates, effectively implementing the QFT-based diffusion operator. This amplitude amplification selectively enhances valid SAWs while suppressing invalid configurations. The complete simulation for N=71 steps computes the enumeration in 26.92 minutes, as summarized in Table~I.

\begin{table}[htbp]
\centering
\caption{Computational time for the enumeration of Self-Avoiding Walks (SAW) on a 2D lattice through quantum algorithm from $N = 41$ to $71$}
\label{tab:saw_data}
\begin{tabular}{c r S[table-format=5.2]}
\hline
$N$ & {Number of SAW} & {Computation time (\si{\second})} \\
\hline
41 & 800381032599158340 & 541.81 \\
43 & 5659667057165209612 & 595.57 \\
45 & 39992704986620915140 & 651.89 \\
47 & 282417882500511560972 & 710.73 \\
49 & 1993185460468062845836 & 772.12 \\
51 & 14059415980606050644844 & 836.07 \\
53 & 99121668912462180162908 & 902.56 \\
55 & 698501700277581954674604 & 971.59 \\
57 & 4920146075313000860596140 & 1043.14 \\
59 & 34642792634590824499672196 & 1117.24 \\
61 & 243828023293849420839513468 & 1193.90 \\
63 & 1715538780705298093042635884 & 1273.08 \\
65 & 12066271136346725726547810652 & 1354.83 \\
67 & 84841788997462209800131419244 & 1439.10 \\
69 & 596373847126147985434982575724 & 1525.93 \\
71 & 4190893020903935054619120005916 & 1615.29 \\
\hline
\end{tabular}
\end{table}
A comparative analysis of execution time between the proposed approach and the classical transfer-matrix method \cite{jensen2013} is carried out for step lengths ranging from N=41 to N=71. The corresponding results are presented in the detailed data table (Table~\ref{tab:saw_data}), and the computational performance of both classical and quantum approaches for self-avoiding walks (SAWs) is systematically assessed.
This specific range was chosen to align with the comparison analysis of precomputed classical data. The sourced classical data established from transfer-matrix algorithms,\cite{jensen2013},  provides computation time in core-hour with varying numbers of processors.\par
\begin{figure}[!htbp]
    \centering
    \includegraphics[width=1.0\linewidth]{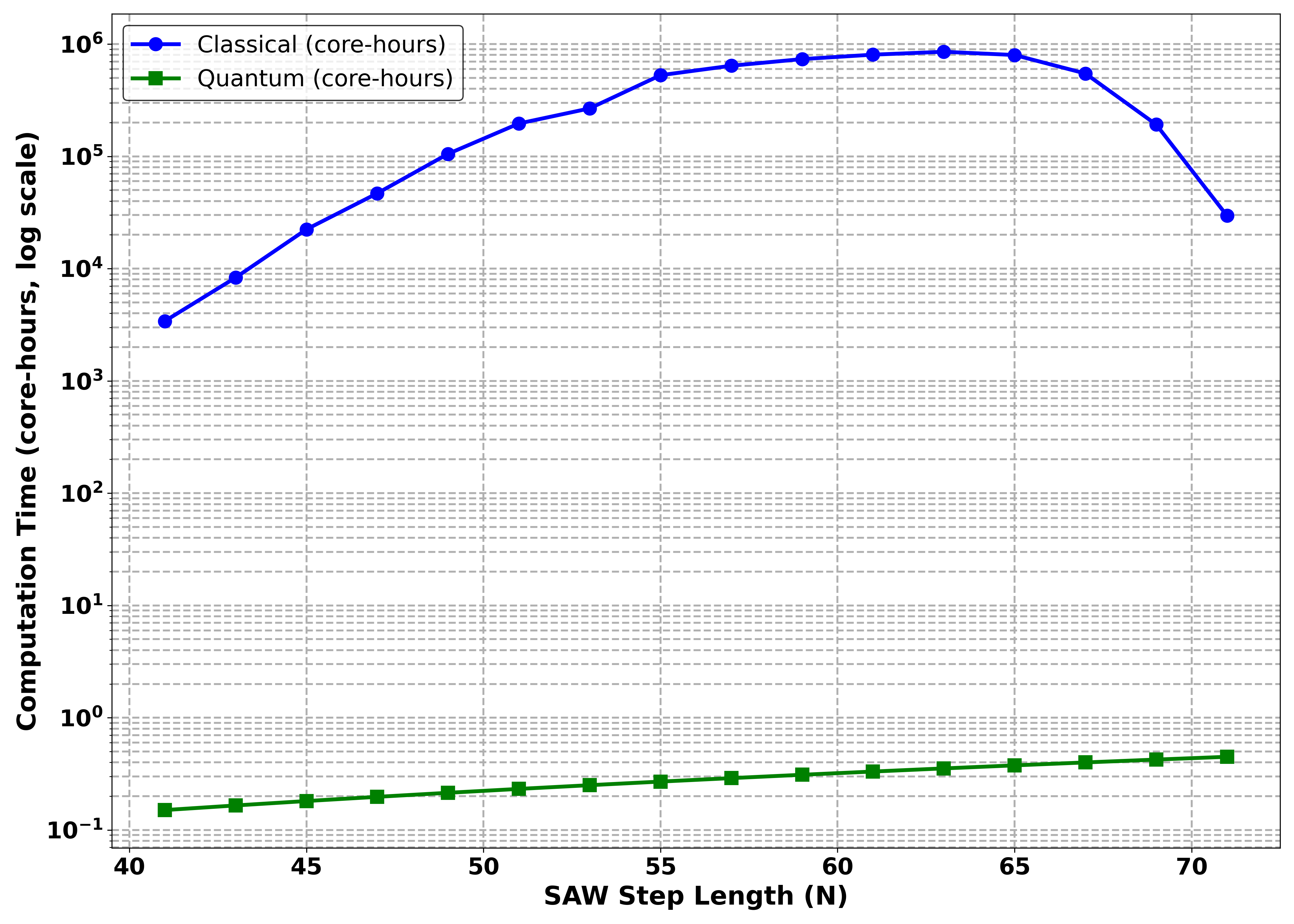} % Adjust width as needed
    \caption{Comparison of computation times for 2D SAW enumeration, showing classical and quantum time in core-hours on a logarithmic scale across $N = 41$ to $71$}
    \label{fig:6}
\end{figure}

In classical algorithms, the computation required range of processor ranges from 48 to 400, due to exponential growth in the computation of possible walks for a given N. Refering to classical data \cite{jensen2013} in Figure~\ref{fig:6}, there was combinatorial growth upto N = 63 with 400 processors in 2,134 hours. Beyond this peak, the CPU time decreases slightly due to the optimization in processor allocation, dropping for N = 71 with 128 processors in 231 hours.

The quantum based computation time was recorded in 1615.29 seconds with 6-physical-cores processor for the complete length up to N = 71 executed on Qiskit (version 2.0.0). From table [\ref{tab:saw_data}] for N = 41, the quantum algorithm took 541.81 seconds $\approx 9.03$ minutes. The time increases steadily with the increase in N, depicting the linear growth in circuit depth and the number of qubits. For N = 55, the execution time is 971.59 seconds $\approx 16.19$ minutes, and by N = 71, it reaches 1615.29 seconds $\approx 26.92$ minutes. This linear growth in time is evident, as the quantum algorithm's complexity scales with the number of steps, and sampling shots $(2^{20})$ remain efficient with Grover iterations $\sqrt{\mu^{N-71}}$.

\subsection{3D SAWs Computational Analysis}
\begin{figure}[!htbp]
    \centering
    \includegraphics[width=1.0\linewidth]{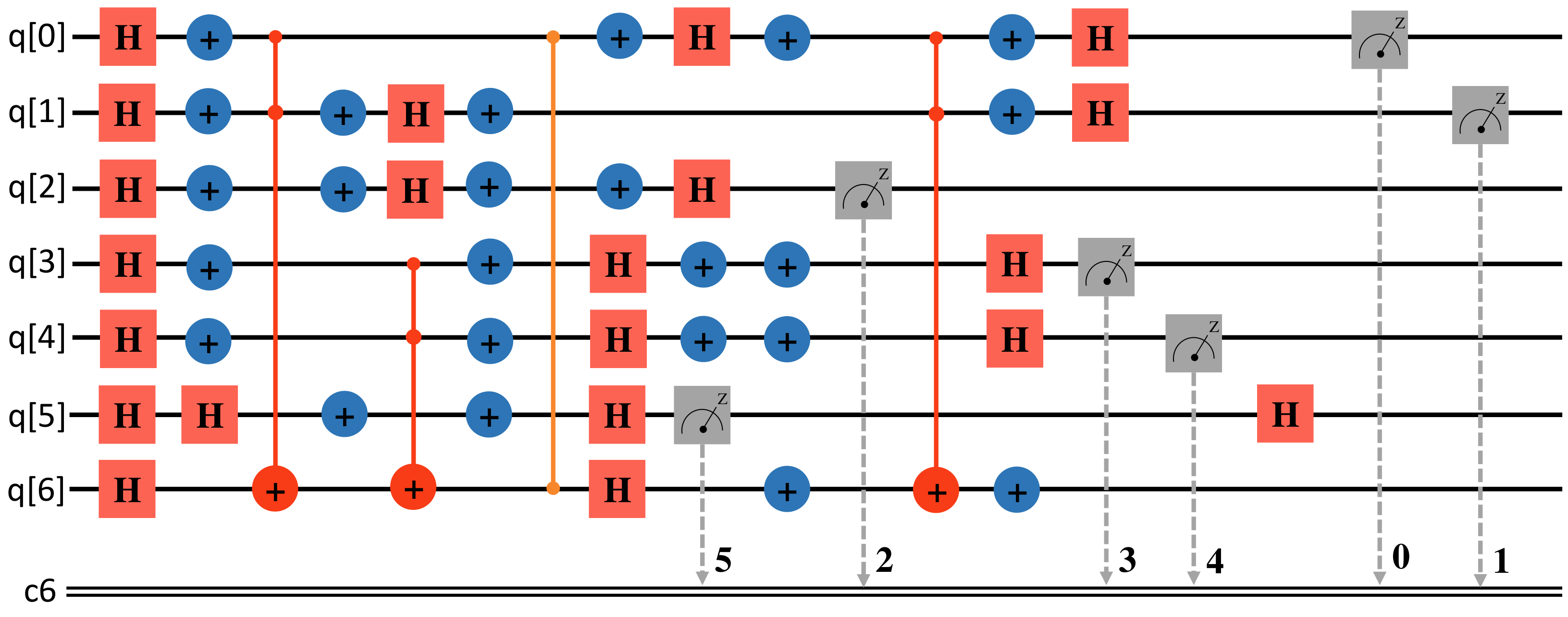}
    \caption{Quantum Circuit demonstration for N=2 in 2D Lattice. Red boxes $(H)$ represent Hadamard Gates for initializing superposition. Blues $(+)$ indicates NOT $X$ gate for phase flip. Red $(+)$ symbols denote the Toffoli $(CCX)$ gate for controlled condition. The yellow line represents the Controlled-Z $(CZ)$ gate, which introduces a phase shift. Grey boxes represent the Z-basis for the measurement at the end of the computation.}
    \label{fig:7}
\end{figure}

The enumeration of SAWs in a 3D lattice requires $(3N+1)$ number of qubits. In Figure~[\ref{fig:7}] we show quantum circuit for 2-step walk in 3D lattice, where ($q[0]$ to $q[2]$) represents $1^{st}$ step and ($q[3]$ to $q[5]$) for $2^{nd}$ step (each step has 3 bits for classical bitstring), and one ancilla qubit $q[6]$ helps to mark the valid SAWs. Further on, Hadamard Gate $(H)$ applied at each qubit from ($q[0]$ to $q[6]$) for complete superposition over all possible walks which quantify the total $64$ configuration with valid and invalid walks. To mark the valid walks, a Quantum oracle implements at each direction to verify backtracking (e.g., moving +X then -X $\rightarrow$ invalid). In the circuit, there is NOT gate $(X)$ which is used to control the marking scheme. This process of marking valid and invalid SAWs is continued through controlled gates: Toffoli gate used for the detection of invalid SAWs and the Controlled-Z gate, which flips the phase of valid SAWs and reinforces the constraints of SAWs. And again, Not gate $(X)$ is implemented to reset the memory qubit to make further valid SAWs. 
As explained in Section~\ref{sec:QAE Impl}, the diffusion operator responsible for suppression of invalid walks via Grover's iteration and $QFT^\dagger$. The execution of a $N=2$ circuit gives the enumeration $Z_N=30$. Further, the algorithm computes the enumeration for $N=40$ in $13.05$ minutes shown in the Table [\ref{tab:3Dsaw_data}]. \\
The Quantum Algorithm (QAE) demonstrate a dramatic reduction in computation time, where as the classical length doubling method by Schram \cite{Schram2011} computed the enumeration upto $Z_{36}$, using SARA supercomputer in Amsterdam it took 50,000 CPU hours (which is equivalent to 250 hours of wall-clock time). Moreover, Schram provided another techniques SAWdoubler \cite{schram2012} to count SAW on a dual-core PC, computed $Z_{28}$ in 1 hour and 40 minutes ($\approx$ {6000} seconds).\\
\begin{figure}[!htbp]
    \centering
    \includegraphics[width=1.0\linewidth]{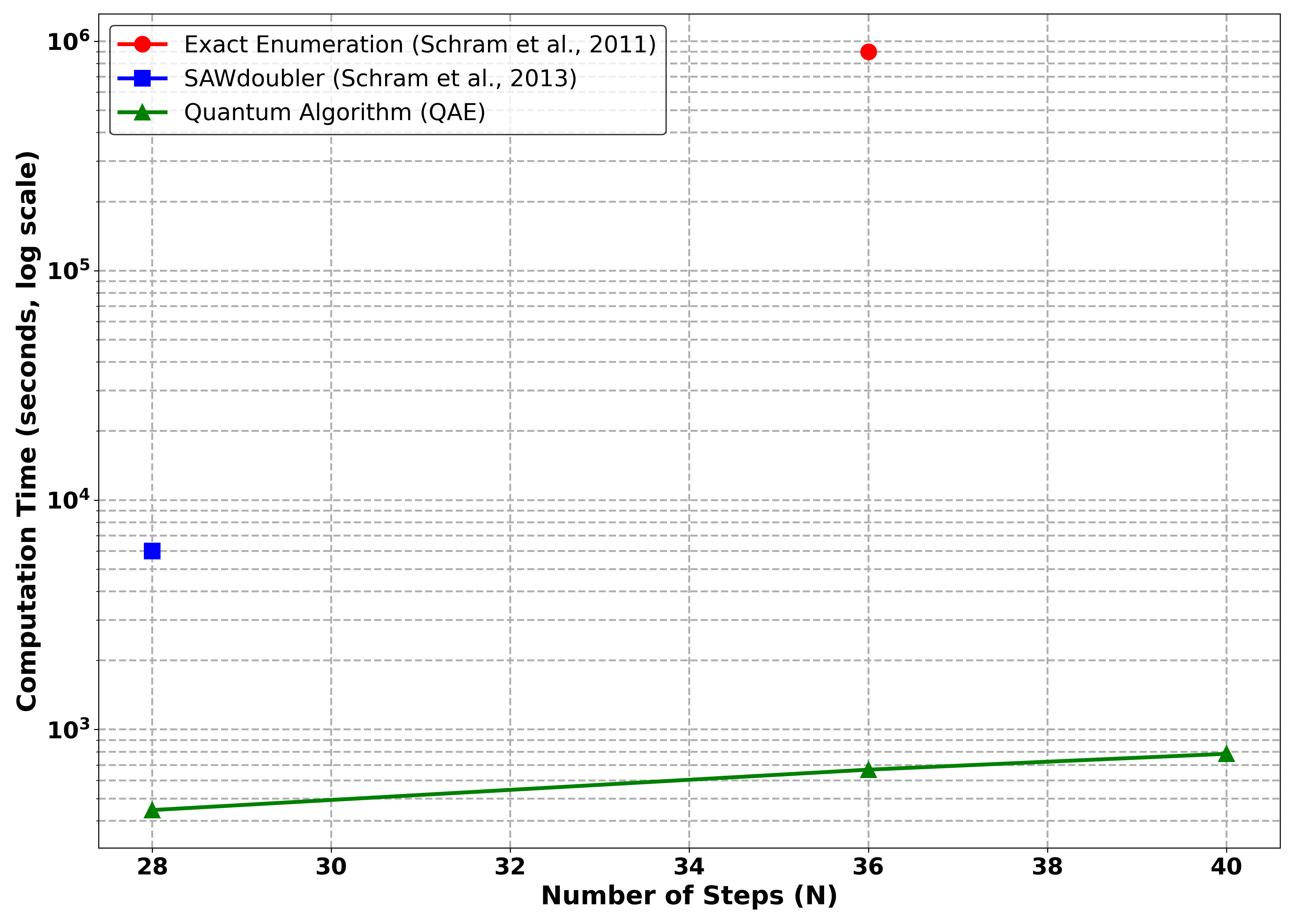} % Adjust width as needed
    \caption{Comparison of computation times for 3D SAW enumeration, showing classical and quantum computation time for $N=28, 36 \& 40$}
    \label{fig:8}
\end{figure}
Whereas the quantum algorithm, took 668.26 seconds ($\approx 11.13$ minutes) to compute the enumeration for \(N=36\), through QAE that matches with classical enumeration; \(Z_{36} = 2941370856334701726560670\) shown in Table [\ref{tab:3Dsaw_data}]. Similarly, for \(Z_{28} = 12198184788179866902\) in 445.45 seconds ($\approx$ 7.42 minutes) shown in Table [\ref{tab:3Dsaw_data}].
These runtimes efficiently demonstrate simulation on current hardware, offers potential advantage over classical transfer-matrix and length-doubling method, which required thousands of CPU-hours for on pre-2020 systems. However, with the advanced classical computing it may still requires hours or days for the computation up to \(N=36\) \cite{Clisby2017}. While full empirical simulation remain infeasible on classical systems, instead shift focus on theoretical asymptotic improvements, as explored in \cite{Zbarsky2019}. Overall, our quantum inspired approach demonstrates promising advantages over classical computing. QAE outperforms the previously computed length and demonstrated enumeration for \(Z_{40}\) on a 3D cubic lattice in a total cumulative time of 783.63 seconds ($\ approx$ 13.05 minutes) [\ref{tab:3Dsaw_data}].\\

\begin{table}[htbp]
\centering
\caption{Computational time for the enumeration of Self-Avoiding Walks (SAW) on a 3D simple cubic lattice through quantum algorithm for $N = 40$; Total time taken to compute: 783.63 seconds}
\label{tab:3Dsaw_data}
\begin{tabular}{c r S[table-format=5.2]}
\hline
$N$ & {Number of SAW} & {Computation time (\si{\second})} \\
\hline
1  & 6                           & 0.04 \\
2  & 30                          & 0.14 \\
3  & 150                         & 0.32 \\
4  & 726                         & 0.56 \\
5  & 3534                        & 0.88 \\
6  & 16926                       & 1.27 \\
7  & 81390                       & 1.73 \\
8  & 387966                      & 2.25 \\
9  & 1853886                     & 2.85 \\
10 & 8809878                     & 3.52 \\
11 & 41934150                    & 4.26 \\
12 & 198842742                   & 5.07 \\
13 & 943974510                   & 5.95 \\
14 & 4468911678                  & 6.91 \\
15 & 21175146054                 & 7.93 \\
16 & 100121875974                & 9.02 \\
17 & 473730252102                & 10.18 \\
18 & 2237723684094               & 11.41 \\
19 & 10576033219614              & 12.72 \\
20 & 49917327838734              & 14.09 \\
21 & 235710090502158             & 15.54 \\
22 & 1111781983442406            & 17.05 \\
23 & 5245988215191414            & 18.64 \\
24 & 24730180885580790           & 20.29 \\
25 & 116618841700433358          & 22.02 \\
26 & 549493796867100942          & 23.82 \\
27 & 2589874864863200574         & 25.68 \\
28 & 12198184788179866902        & 27.62 \\
29 & 57466913094951837030        & 29.63 \\
30 & 270569905525454674614       & 31.71 \\
31 & 1274191064726416905966      & 33.86 \\
32 & 5997359460809616886494      & 36.08 \\
33 & 28233744272563685150118     & 38.37 \\
34 & 132853629626823234210582    & 40.73 \\
35 & 625248129452557974777990    & 43.16 \\
36 & 2941370856334701726560670   & 45.66 \\
37 & 13837797706226707233505280  & 48.23 \\
38 & 65092880261421794634760192  & 50.87 \\
39 & 306162469755604615651393536 & 53.59 \\
40 & 1439875035115056625591779328 & 56.35 \\
\hline
\end{tabular}
\end{table}

\section{Conclusion}
 In this paper, we have developed a quantum computing inspired algorithm to count the number of self-avoiding walks (SAWs) on lattices in 2D and 3D. The QAE approach exhibit a significant advantage in computational time over classical methods, particularly for longer walk lengths up to length $N=40$ in $13.06$ minutes and $N=71$ in $26.92$ minutes for 2D and 3D lattices, respectively. Results show how quantum computing can be used to solve combinatorial time complexity issues in polymer physics and statistical mechanics. The enhanced efficiency achieved illustrates not only the growing maturity of quantum algorithms but also their practical applicability in domains where classical enumeration becomes computationally prohibitive. Overall, this work opens avenues for further quantum-based investigations of complex lattice path problems and other NP-hard combinatorial challenges.
% This work paves the way for further quantum-based investigations into complex lattice path problems and other NP-hard combinatorial challenges.\\

 \begin{acknowledgments}
% ... Other acknowledgments ...

The source code implementing the quantum SAW enumeration algorithm for 2D/3D lattices and simulation scripts, is publicly available at \url{https://github.com/with-hemant/Quantum-amplitude-estimation.git}.

\end{acknowledgments}

\par
\appendix 

\section*{Appendix}
\label{app:supplement}  

\setcounter{equation}{0}
\setcounter{figure}{0}
\setcounter{table}{0}

\section{Quantum Superposition for the given input length N}
\label{app:superposition}

For uniform superposition in quantum circuit. Each steps require $(2N)$ qubits, where $2N$ represents number of qubits with input length N (For example; If $N =1$ then the qubits count will be $(2)$ encoding 4 direction),
To achieve uniform superposition, Hadamard gates(H) applied to the 2N qubit;
\begin{equation}
    H \lvert 0 \rangle = \frac{\lvert 0 \rangle + \lvert 1 \rangle}{\sqrt{2}}
\end{equation}

\begin{align}
H \otimes H \ket{00} 
    &= \left( \frac{\ket{0} + \ket{1}}{\sqrt{2}} \right) 
       \otimes \left( \frac{\ket{0} + \ket{1}}{\sqrt{2}} \right) \nonumber \\[8pt]
    &= \frac{ \ket{00} + \ket{01} + \ket{10} + \ket{11} }{2}
\end{align}

it create superposition over all $2^{2N}$ in possible bitstring. After applying H to each of 2N qubits the initial state $|0\rangle^{\otimes 2N}$ to account valid and invalid walks:
\begin{equation}
    |\psi{_{0}}\rangle = \frac{1}{\sqrt{4^{N}}} \sum_{i=0}^{2^{2N}-1} \alpha_i |i\rangle
    \label{eq:i.sup}
\end{equation}

where, $|i\rangle$ represent the set of sequences corresponding to the basis state for specific walk configuration in 2D  and $\alpha_i$ is  the amplitudes of basis states. 2N qubits are grouped into N pairs, each pair represents one step direction. The 2-qubit pair ($|00\rangle, |01\rangle, |10\rangle, |11\rangle)$ corresponding to the 4 direction ($i\in{0,1,2,3}$). \\
\begin{enumerate}
\item \textbf{Quantum Encoding}
\begin{itemize}
\item $|00\rangle \rightarrow right(+X)$
\item $|01\rangle \rightarrow right(-X)$
\item $|10\rangle \rightarrow right(+Y)$
\item $|11\rangle \rightarrow right(-Y)$
\end{itemize}

\item \textbf{Total Choices at each steps} $\rightarrow 4^{N}$

\item \textbf{Nomalization:} Putting the value of Amplitude $\alpha_i$ = $\frac{1}{\sqrt{4^N}}$ in Equation~\ref{eq:i.sup} \\
\[
\sum_{i=0}^{2^{2N}-1} |\alpha_i|^2 = 1 
\]
\begin{equation}
    \sum_{i=0}^{2^{2N}-1} |\frac{1}{\sqrt{4^N}}|^2 = 4^{N} \cdot \frac{1}{4^{N}} = 1 
\end{equation}
\end{enumerate}
\subsection{For N = 1 SAWs in 2D Lattice}
\begin{enumerate}
\item \textbf{Basis States}
\begin{itemize}
\item $|00\rangle (decimal; |0\rangle) \rightarrow right(+X)$
\item $|01\rangle (decimal; |1\rangle) \rightarrow right(-X)$
\item $|10\rangle (decimal; |2\rangle) \rightarrow right(+Y)$
\item $|11\rangle (decimal; |3\rangle) \rightarrow right(-Y)$
\end{itemize}
\item \textbf{Amplitude:} For N = 1, $4^{1} = 4$
\begin{equation}
    \frac{1}{\sqrt{4}} = \frac{1}{2}
\end{equation}

\item \textbf{Initial State:} The superposition state for step 1; According to equation ~\eqref{eq:i.sup}
\[
|\psi_{0}\rangle =\frac{1}{\sqrt{4^N}} \sum_{i=0}^{2^{2N}-1} |i\rangle
\]
\begin{equation}
    |\psi_{0}\rangle =\frac{1}{2} \sum_{i=0}^{3} |i\rangle
\end{equation}
\[
|\psi_0\rangle = \frac{1}{2} \left( |00\rangle + |01\rangle + |10\rangle + |11\rangle \right)
\]
or in decimal notation:
\begin{equation}
    |\psi_0\rangle = \frac{1}{2} \left( |0\rangle + |1\rangle + |2\rangle + |3\rangle \right)
\end{equation}

\item \textbf{Nomalization:}
\[
\alpha_i = \frac{1}{2} for i \in \{0, 1, 2, 3\} 
\]
\[
4 \cdot \left(\frac{1}{2}\right)^2 = 4 \cdot \frac{1}{4} = 1
\]
\end{enumerate}
Above normalization quantifies that all the probable walk for step 1 is valid.
\subsection{For N = 2 SAWs in 2D Lattice}
\begin{enumerate}
\item \textbf{Basis States:} Each sequence is encoded by 4-qubits it consist of $|q_1 q_2 q_3 q_4\rangle$ where $(q_1 q_2)$ represented for first step and $(q_1 q_2)$ for second step.
\begin{itemize}
\item $ |0000\rangle = |0\rangle $: Right, Right
\item $ |0001\rangle = |1\rangle $: Right, Left
\item $ |0010\rangle = |2\rangle $: Right, Up
\item $ |0011\rangle = |3\rangle $: Right, Down
\item $ |0100\rangle = |4\rangle $: Left, Right
\item $ |0101\rangle = |5\rangle $: Left, Left
\item $ |0110\rangle = |6\rangle $: Left, Up
\item $ |0111\rangle = |7\rangle $: Left, Down
\item $ |1000\rangle = |8\rangle $: Up, Right
\item $ |1001\rangle = |9\rangle $: Up, Left
\item $ |1010\rangle = |10\rangle $: Up, Up
\item $ |1011\rangle = |11\rangle $: Up, Down
\item $ |1100\rangle = |12\rangle $: Down, Right
\item $ |1101\rangle = |13\rangle $: Down, Left
\item $ |1110\rangle = |14\rangle $: Down, Up
\item $ |1111\rangle = |15\rangle $: Down, Down
\end{itemize}
\item \textbf{Amplitude:} For N = 1, $4^{2} = 16$
\[
so, \frac{1}{\sqrt{16}} = \frac{1}{4}
\]

\item \textbf{Initial State:}
\[
|\psi_{0}\rangle =\frac{1}{\sqrt{4^N}} \sum_{i=0}^{2^{2N}-1} |i\rangle
\]
\begin{equation}
    |\psi_{0}\rangle =\frac{1}{\sqrt{4^2}} \sum_{i=0}^{15} |i\rangle
\end{equation}
\begin{multline*}
|\psi_0\rangle = \frac{1}{4} (|0000\rangle + |0001\rangle + |0010\rangle + |0011\rangle + |0100\rangle + |0101\rangle + |0110\rangle\\
 + |0111\rangle + |1000\rangle + |1001\rangle + |1010\rangle + |1011\rangle \\ 
+ |1100\rangle + |1101\rangle + |1110\rangle + |1111\rangle)
\end{multline*}
\item \textbf{SAW Condition:} For 2-step walker occupies 3 - lattice point.
\begin{itemize}
\item Lattice point after first step $\rightarrow P_{1}$ 
\item Lattice point after second step $\rightarrow P_{2}$ 
\end{itemize} 
Ideally quantifies as;
\begin{itemize}
\item First step $(q_1 q_2)$: Move from (0,0) to $P_{1}$
\item Second step $(q_3 q_4)$: Move from $P_{1}$ to $P_{2}$
\end{itemize} 

\item \textbf{Nomalization:}
\begin{equation}
    \sum_{i=0}^{15} \left| \frac{1}{4} \right|^2 = 16 \cdot \frac{1}{16} = 1
    \label{eq:norm-2}
\end{equation}
\end{enumerate}
The Normalization of equation~\eqref{eq:norm-2} signifies that Quantum state is valid. Applying the key Quantum algorithm, \textbf{QAE} encoded with Grover's Search amplify the amplitude of valid SAWs and suppress the amplitude of Invalid SAWs.
\begin{equation}
    |\psi\rangle = \frac{1}{\sqrt{12}} \sum_{i \in \{0,2,3,5,6,7,8,9,10,12,13,15\}} |i\rangle
\end{equation}
Again normalization check to validate the observed valid SAWs; $12 \cdot \left( \frac{1}{\sqrt{12}} \right)^2 = 12 \cdot \frac{1}{12} = 1$.\\

%==========================================================
\begin{center}
\section{Derivation of $\alpha_i$ for 2D and 3D lattice}
\label{app:amplitude}
\end{center}
\begin{itemize}
\item \textbf{Number of states:} In 2D lattice there are $4^N$ possible sequences, each N steps has 4 possible direction. We are using 2-qubits at each step, $2^{2N}$ = $4^N$ this quantifies all possible direction sequences.
\item \textbf{Uniform Superposition:} For creating uniform superposition for $4^N$ states, where each basis states $|i\rangle$ encoded with amplitude $\alpha_i$.
\begin{equation}
    \sum_{i=0}^{2^{2N}-1} |\alpha_i|^2 = 1
\end{equation}
\begin{equation}
    4^N \cdot |\alpha_i|^2 = 1
    \label{eq:alpha1}
\end{equation}
\[
|\alpha_i|^2 = \frac{1}{4^N}
\]
\begin{equation}
    \alpha_i = \frac{1}{\sqrt{4^N}}
\end{equation}
Since the interpretation of the amplitude will ensure equal probability;
\begin{equation}
    |\alpha_i|^2 = \left( \frac{1}{\sqrt{4^N}} \right)^2 = \frac{1}{4^N}
    \label{eq:alpha2}
\end{equation}
From the equation~\eqref{eq:alpha1} and ~\eqref{eq:alpha2}
\[
4^N \cdot \frac{1}{4^N} = 1
\]
Thus, $ \alpha_i = \frac{1}{\sqrt{4^N}} $ quantifies the valid amplitude.
\end{itemize}

\begin{itemize}
\item \textbf{Number of states:} In 3D lattice there are $6^N$ possible sequences, each N steps has 6 possible direction. We are using 3-qubits at each step, $2^{3N}$ = $6^N$ this quantifies all possible direction sequences.
\item \textbf{Uniform Superposition:} For creating uniform superposition for $6^N$ states, where each basis states $|i\rangle$ encoded with amplitude $\alpha_i$.
\begin{equation}
    \sum_{i=0}^{2^{3N}-1} \left| \frac{\alpha_i}{\sqrt{6^N}} \right|^2 = 1
\end{equation}
The set of indices for valid direction sequences can be represented as $V$. For $ i \in V $, set $\alpha_i = 1$ and for $ i \notin V $ sets  $\alpha_i = 0$
\begin{equation}
    \sum_{i \in V} \left| \frac{1}{\sqrt{6^N}} \right|^2 = 6^N \cdot \frac{1}{6^N} = 1
\end{equation}
For invalid sequences ($ i \notin V $):
\begin{equation}
    \sum_{i \notin V} \left| \frac{0}{\sqrt{6^N}} \right|^2 = 0
\end{equation}
Thus, $ \alpha_i = \frac{1}{\sqrt{6^N}} $ quantifies the valid amplitude with $i \in V$.
\end{itemize}

%==========================================================================
\begin{center}
\section{QAE Derivation}
\label{app:QAE}
\end{center}
The algorithm QAE provides an efficient way for enumeration SAWs, which consist of some essential components that are;
\begin{itemize}
\item \textbf{Walk Register:} The algorithm uses a quantum register to represent the walk on a lattice, where 2-qubits and 3-qubits are used for each step corresponding to a 2D and 3D lattice.
\item \textbf{Memory Register:} A quantum register which account the visited lattice site to enforce self-avoidance.
\item \textbf{Flag Qubit:} In a single qubit makring the valid walk $|1\rangle$ else invalid walk  $|0\rangle$
\end{itemize}
The total state space (i.e., Hilbert Space)
\begin{equation}
    \mathcal{H} = \mathcal{H}_{\text{walk}} \otimes \mathcal{H}_{\text{memory}} \otimes \mathcal{H}_{\text{flag}}
\end{equation}
where:
\begin{itemize}
\item \textbf{$\mathcal{H}_{\text{walk}}$}: 2N qubits space for possible walk direction ($4^N$)
\item \textbf{$\mathcal{H}_{\text{memory}}$}: Accounts for overall space to trace the visited sites.
\item \textbf{$\mathcal{H}_{\text{flag}}$}: Confirms the valid SAWs.23
\end{itemize}
Initial state of all qubits $|0\rangle^{\otimes k}$, where $k$ represents total number of qubits.
\begin{enumerate}
    \item \textbf{Constructing Quantum Operator A} It must prepare superposition over all possible walks and then confirms the valid SAWs through flag qubit.
    \begin{enumerate}[label=\Roman*.]
        \item \textbf{Step 1: Superposition Walk Generation} Initializes with walk register by applying Hadamard gate $H$ to each qubits.
        \begin{equation}
            H^{\otimes 2n} |0\rangle^{\otimes 2n} = \frac{1}{\sqrt{4^N}} \sum_{w \in \{0,1,2,3\}^N} |w\rangle,
        \end{equation}
        where, $w$ = $(d_{1},.....,d_{n})$ shows walk with $d_{i} \in \{0, 1, 2, 3\}$.In which memory and flag registers are encoded $|0\rangle_{\text{memory}} |0\rangle_{\text{flag}}$, so the state will be
        \begin{equation}
            \frac{1}{\sqrt{4^N}} \sum_{w \in \{0,1,2,3\}^N} |w\rangle |0\rangle_{\text{memory}} |0\rangle_{\text{flag}}
        \end{equation}
    \item \textbf{Step 2: Track Visited Sites} Enforcing self-avoidance, to the visited sites and defining a unitary $U_{path}$ that maps;
    \begin{equation}
        |w\rangle|0\rangle_{memory} \rightarrow |w\rangle|path(w)\rangle
    \end{equation}
    where, $|path(w)\rangle$ represents sequence of visited site by walk $w$
    \end{enumerate}
    % Assuming this is inside an outer enumerate (from your code: \begin{enumerate} ... \item Step 1 ... \item Step 2 ... then this)
\item \textbf{Step 3: Self-Avoidance Check} Quantum oracle applied $U_{\text{SAW}}$ to verify the constraint by examining $|\text{path}(w)\rangle$. The oracle uses flag qubit to $|1\rangle$ for valid SAWs.
    \begin{equation}
    \begin{split}
        U_{\text{SAW}} |w\rangle &\, |\text{path}(w)\rangle |0\rangle_{\text{flag}} \\
        &= |w\rangle |\text{path}(w)\rangle |f(w)\rangle_{\text{flag}}
    \end{split}
    \label{eq:usaw}
    \end{equation}
    where $|f(w)\rangle = |1\rangle$ for valid $w$ and $|f(w)\rangle = |0\rangle$ for invalid ones.

\item \textbf{Step 4: Quantum Operator $A$}
    \begin{equation}
    \begin{split}
        A &= (U_{\text{SAW}} \otimes I_{\text{flag}}) \\
          &\quad (U_{\text{path}} \otimes I_{\text{flag}}) \\
          &\quad (H^{\otimes 2n} \otimes I_{\text{memory}} \otimes I_{\text{flag}})
    \end{split}
    \label{eq:operator-a}
    \end{equation}

    % Use description for sub-steps: Better alignment for labels + inline math
    \begin{description}[leftmargin=*,labelwidth=2cm]  % Compact, fixed label width
    \item[Hadamard Process:] 
        {\small
        \begin{equation*}
        \frac{1}{\sqrt{4^N}} \sum_w |w\rangle |0\rangle_{\text{memory}} |0\rangle_{\text{flag}}
        \end{equation*}
        }
    \item[Path Computation Process:] 
        {\small
        \begin{equation*}
        \begin{split}
            \frac{1}{\sqrt{4^N}} &\sum_{w} |w\rangle \\
            &\quad |\text{path}(w)\rangle |0\rangle_{\text{flag}}
        \end{split}
        \end{equation*}
        }
    \item[Quantum Oracle:] 
        {\small
        \begin{equation*}
        \begin{split}
            \frac{1}{\sqrt{4^N}} &\sum_{w} |w\rangle \\
            &\quad |\text{path}(w)\rangle |f(w)\rangle_{\text{flag}}
        \end{split}
        \end{equation*}
        }
    \end{description}

    % Use description for valid/invalid: Consistent with above, bold labels
    \begin{description}[leftmargin=*,labelwidth=2.5cm]
    \item[\textbf{Valid SAWs}:] There will be $Z_N$ walks then $f(w)=1$.
    \item[\textbf{Invalid SAWs}:] There will be $4^N - Z_N$ walks then $f(w)=0$.
    \end{description}

    Combining all the mentioned steps to form:
    \begin{equation}
    \begin{split}
        A|0\rangle &= \frac{1}{\sqrt{4^N}} \sum_{w: f(w)=1} 
                      |w\rangle |\text{path}(w)\rangle |1\rangle \\[4pt]
        &+ \frac{1}{\sqrt{4^N}} \sum_{w: f(w)=0} 
            |w\rangle |\text{path}(w)\rangle |0\rangle
    \end{split}
    \label{eq:a-zero}
    \end{equation}
\end{enumerate}
The QAE equation can be rewrite as; 
\begin{equation}
    A|0\rangle = \sqrt{a}|1\rangle|\psi_1\rangle + \sqrt{1-a}|0\rangle|\psi_0\rangle
\end{equation}
where,
\begin{itemize}
\item \textbf{A:} A quantum operator that produces the complete superposition and execute for the resulting state.
\item \textbf{$|1\rangle|\psi_1\rangle$}: Produces ``true'' state corresponding to valid SAWs.
\item \textbf{$|0\rangle|\psi_0\rangle$}: Produces ``false'' state corresponding to invalid SAWs.
\item \textbf{a}: represents the probability of generating valid SAWs 
\[
\sqrt{a} = \sqrt{\frac{Z_N}{M}}
\]
\begin{equation}
    \sqrt{a} = \sqrt{\frac{Z_N}{4^N}} 
\end{equation}
where, $Z_N$ number of SAWs of length N. and $4^N$ shows total number of possible walks
\item \textbf{$\sqrt{1-a}$}: represents the probability of generating invalid SAWs, since the value of a;
\[
a=\frac{Z_N}{4^N}
\]
\begin{equation}
    \sqrt{1-a} = \sqrt{\frac{4^N - Z_N}{4^N}}
\end{equation}
\end{itemize}
The ``True" state;
\begin{equation}
    |1\rangle |\psi_1\rangle = \frac{1}{\sqrt{Z_N}} \sum_{w: f(w)=1}|w\rangle |\text{path}(w)\rangle |1\rangle
\end{equation}
where $|\psi_1\rangle = \frac{1}{\sqrt{Z_N}} \sum_{w: f(w)=1} |w\rangle |\text{path}(w)\rangle$ is normalized state that quantify valid SAWs.\\
The ``False" state;
\begin{equation}
    |0\rangle |\psi_0\rangle = \frac{1}{\sqrt{4^N - Z_N}} \sum_{w: f(w)=0} |w\rangle |\text{path}(w)\rangle |0\rangle
\end{equation}
where, $|\psi_0\rangle = \frac{1}{\sqrt{4^N - Z_N}} \sum_{w: f(w)=0} |w\rangle |\text{path}(w)\rangle |0\rangle$ quantify invalid SAWs.\\
Thus, the QAE equation will be;
\[
A|0\rangle = \sqrt{a} |1\rangle |\psi_1\rangle + \sqrt{1-a} |0\rangle |\psi_0\rangle
\]

%==============--===================================
\begin{center}
\section{Computation for bigger steps in 3D Simple Cubic Lattice}
\label{app:computation}
\end{center}
In the QAE algorithm, A key class named "QuantumSAWCounter3D" is constructed to provide logic for counting 3D SAWs, which uses Qiskit library to simulate the quantum circuit upto \(N=40\). The circuit utilizes ;
\begin{equation}
    n = 3\times(N-36) + 1
\end{equation}
It has similar implementation to propel uniform superposition over each steps mentioned earlier was \(2N+1\) in which $2N$ represent 2 qubits per steps with number of step. Here,  $3\cdot(N-36)$ encodes the direction for additional step enumeration (3 qubits per step to represent available direction) and plus one qubits represents ancilla qubit to be used in QAE amplification process.
\begin{itemize}\tightlist
\item{Hadamard gates $(H)$ applied to create uniform superposition over all possible walks}.
\item{Multi-controlled gates $(MCX)$ used to encode the constraints for the avoidance of intersection and provide valid transitions.}.
\item{Grover's operator implemented with $\text{QFT}^\dagger$ (Inverse QFT) where the controlled-Z, Hadamard and multi-controlled X gates to amplify the valid states with iteration optimization $\sqrt{{\mu}^{(N-36)}}$}.
\item{The overall measurement will collapse in form of bitstring to represent possible SAWs enumeration.}
\end{itemize}
The measurement take place once the bitstring is executed the \(valid\_extension\) will verify avoidance of intersection and no backtracking with \(N-36\) steps. Fraction to calculate valid bitstring in the outcome; $\frac{totalvalid}{shots}$ Each group of 3-qubits encodes direction (from 0 to 5), mapped in accordance with paths +X, -X, +Y, -Y, +Z, -Z. The circuit execution take place with shots = $2^{20}$ $(\approx1,048,576)$ times which increases the statistical accuracy of the valid SAWs and reduces random fluctuations from the outcome was chosen to maintain the computational convenience for stable circuit execution without being noisy.\\
The total number of valid bitstrings calculated as;
\begin{equation}
    N-36 steps \approx 2^{3\cdot(N-36)}
\end{equation}
in which each steps encoded with 3 qubits. So, the number of valid SAWs will be represented as;
\begin{equation}
    Z_{N} = \frac{total valid}{shots}\cdot 2^{3\cdot(N-36)}
\end{equation}
According to classical estimation the sampling for large N follows asymptotic growth, to which the QAE presents an asymptotic correction formula it illustrates the enumeration with of given length N;
\begin{equation}
    Z_N = Z_{36} \cdot \left( \mu^{N - 36} \cdot \left( \frac{N}{36} \right)^{\theta} \right)
\end{equation}
where, $\mu$ = 4.684 represents connective constant to account the average number of "valid\_extension", $\theta$ = 0.1597 shows critical exponents to confirm the subleading correction and constant prefactor of lattice $A$ will not be accounted. While analysis the $A$ becomes negligible for large $N$ due to dominant sampling of grover's iteration as the $\mathcal{(O)}$ notation focuses on dominant factor with faster growing terms of valid SAWs. $(\frac{N}{36})^{\theta}$ represents scaling ratio consistent with number of input length $N$ to encode the growth from \(N = 36\) and $\theta$ confirms the subleading correction should be constant. So, the program execute the growth factor;
\begin{equation}
    growth\_factor = \left( \mu^{N - 36} \cdot \left( \frac{N}{36} \right)^{\theta} \right)
\end{equation}
Thus, the equation for the computation of larger step will be;
\begin{equation}
    Z_N = [Z_{36} \cdot growth\_factor]
\end{equation}
\subsection{Example Calculation for N =37}
\begin{flushleft}
\begin{itemize}\tightlist
\item \textbf{Input:} {N = 37, base step = 36}.
\item \textbf{Base Step Value:} {$Z_{36}$ = 2941370856334701726560670}.
\item \textbf{Parameters:}
  \begin{itemize}
    \item $\mu = 4.684$
    \item $\theta = 0.1597$
    \item $N - 36 = 1$
    \item $\dfrac{N}{36} = \dfrac{37}{36} \approx 1.02778$
  \end{itemize}
  
  \item \textbf{Growth Factor:}
  
  \begin{equation}
      \text{growth\_factor} = \mu^{1} \cdot \left( \frac{37}{36} \right)^{\theta}
  = 4.684 \cdot (1.02778)^{0.1597}
  \end{equation}
  \item Compute $(1.02778)^{0.1597} \approx 1.0044$ \\
  \item Using $\ln(1.02778) \approx 0.0274$, so $0.1597 \cdot 0.0274 \approx 0.00438$, and $e^{0.00438} \approx 1.0044$.\\
Thus, the growth factor for $N=40$
\begin{equation}
    \text{growth\_factor} \approx 4.684 \cdot 1.0044 \approx 4.7046
    \label{eq:Enum37}
\end{equation}
  \item \textbf{SAW Count:} Enumeration for $N=37$ steps using the growth factor from equation~\eqref{eq:Enum37}
  \begin{equation}
\begin{split}
    z_{37} &= z_{36} \cdot 4.7046 \\[4pt]
    &\approx 2941370856334701726560670 \cdot 4.7046
\end{split}
\label{eq:z37}  % Optional label; adjust as needed
\end{equation}
  \[
  \approx 1.383779770622670723350528 \times 10^{28}
  \]

  \begin{itemize}
    \item Accounting the estimation for N = 37:
    \[
    z_{37} = 13837797706226707233505280
    \]
  \end{itemize}

  \item \textbf{Output:} Matches the program’s result for $N = 37$.
\end{itemize}
\end{flushleft}

\clearpage

\onecolumngrid

\begin{center}
    \large \textbf{\MakeUppercase{References}}  
\end{center}

\twocolumngrid
\bibliography{references}

\begin{thebibliography}{9}

\bibitem{slade1994}
G. Slade, \textit{Math. Intelligencer} \textbf{16}, 29-35 (1994).
\bibitem{Madras1993}
N. Madras and G. Slade, \textit{The Self-Avoiding Walk} (Springer Science \& Business Media, 2013).
\bibitem{Singh2010}
A.R. Singh, D. Giri, and S. Kumar, \textit{J. Chem. Phys.} \textbf{132}, 235105 (2010).
\bibitem{Singh2009}
A.R. Singh, D. Giri, and S. Kumar, \textit{J. Chem. Phys.} \textbf{131}, 065103 (2009).
\bibitem{ARSingh2009}
A.R. Singh, D. Giri, and S. Kumar, \textit{Phys. Rev. E} \textbf{79}, 051801 (2009).
\bibitem{Zbarsky2019}
S. Zbarsky, \textit{J. Phys. A: Math. Theor.} \textbf{52}, 505001 (2019).
\bibitem{Kesten1963}
H. Kesten, \textit{J. Math Phys.} \textbf{4}, 960 (1963).
\bibitem{Orr1947}
W. J. C. Orr, \textit{Trans. Faraday Soc.} \textbf{43}, 12 (1947).
\bibitem{Fisher1959}
M. E. Fisher and M. F. Sykes, \textit{Phys. Rev.} \textbf{114} 45 (1959).
\bibitem{Guttmann1987}
A. J. Guttmann, \textit{J. Phys. A: Math. Gen.}, \textbf{20}, 1839 (1987).
\bibitem{Guttmann1989}
A. J. Guttmann, \textit{J. Phys. A: Math. Gen.}, \textbf{22}, 2807 (1989).
\bibitem{MacDonald1992}
D. MacDonald, D. L. Hunter, K. Kelly and N. Jan, \textit{J. Phys. A: Math. Gen.} \textbf{25}, 1429 (1992).
\bibitem{MacDonald2000}
D. MacDonald, S. Joseph, D. L. Hunter, L. L. Moseley, N. Jan and A. J. Guttmann, \textit{J. Phys. A: Math. Gen.} \textbf{33}, 5973 (2000).
\bibitem{jensen2013}
I. Jensen, arXiv:1309.6709 (2013).
\bibitem{Schram2011}
R. D. Schram, G. T. Barkema and R. H. Bisseling, \textit{J. Stat. Mech.: Theory Exp.} P06019 (2011).
\bibitem{sutor2019}
R. S. Sutor, \textit{Dancing with Qubits} (Packt Publishing Ltd., 2019).
\bibitem{grinko2021}
D. Grinko, J. Gacon, C. Zoufal and S. Woerner, \textit{npj Quantum Inf.} \textbf{7}, 52 (2021).
\bibitem{Stoudenmire2023}
E. M. Stoudenmire and X. Waintal, arXiv:2303.11317 (2023).
\bibitem{Mandviwalla2018}
A. Mandviwalla, K. Ohshiro and B. Ji, \textit{Proc. IEEE Int. Conf. Big Data}, 2531 (2018).
\bibitem{Kandala2019}
A. Kandala, K. Temme, A.D. Corcoles \textit{et al.}, \textit{Nature} \textbf{567}, 491 (2019).
\bibitem{tensor2023}
S. -X. Zhang, J. Allcock, Z. -Q. Wan \textit{et al.}, \textit{Quantum} \textbf{7}, 912 (2023).
\bibitem{Oracle1994}
A. Berthiaume and G. Brassard, \textit{J. Mod. Opt.} \textbf{41}, 2521 (1994).
\bibitem{pittenger2012}
A. O. Pittenger, \textit{An Introduction to Quantum Computing Algorithms} (Springer Science \& Business Media, 2012).
\bibitem{Sutorpython2019}
R. S. Sutor, \textit{Dancing with Python: Learn to Code with Python and Quantum Computing} (Packt Publishing Ltd., 2021).
\bibitem{Clisby2017}
R. D. Schram, G. T. Barkema, R. H. Bisseling and N. Clisby, \textit{J. Stat. Mech.: Theory Exp.} 083208 (2017).
\bibitem{schram2012}
R. D. Schram, G. T. Barkema and R. H. Bisseling, \textit{Comput. Phys. Commun.} \textbf{184}, 891 (2013).

%\bibitem{kempe2003}
%J Kempe (2003), Quantum Random Walk: An Introductory overview, \textit{Contemporary Physics}, 44:4, 307-327.
%\bibitem{Camilleri2021}
%E. Camilleri, P.R. Rohde, J. Twamley (2021), Self-avoiding quantum walks, \textit{Cornell University}, arXiv:1401.1869v1 [quant-ph]
%\bibitem{Andraca2012}
%Venegas-Andraca, S.E. (2012), Quantum walks: a comprehensive review, \textit{Quantum Inf Process 11, 1015–1106}.
%\bibitem{sun2024}
%Y. Sun, L. Wu (2024), Quantum search algorithm on weighted databases, \textit{Nature, Sci Rep 14, 30169}.
%\bibitem{Kendon2020}
%Viv Kendon (2020), How to compute using Quantum Walks, \textit{Cornell University}, arXiv:2004.01329
%\bibitem{Amit1983}
%D.J. Amit, G. Parisi, L. Peliti (1983), Asymptotic behaviour of the "true" self avoiding walk, \textit{Physiscal Review B.27.1635}
%\bibitem{McGoech2024}
%C. McGeoch (2024), How NOT to fool the Masses when giving performance results for Quantum Computers, \textit{arXiv:2411.08860v1 [quant-ph]}
%\bibitem{Micheletti2021}
%C. Micheletti, P.Hauke and P. Faccioli (2021), Polymer Physics by Quantum Computing, \textit{PHYSICAL REVIEW LETTERS 127, 080501}
%\bibitem{Bepari2022}
%K. Bepari, S. Malik, M. Spannowsky and S. Williams (2021), Quantum walk approach to simulating parton showers, \textit{PHYSICAL REVIEW D 106}
%\bibitem{Rathore2025}
%O. Rathore, A. Basden, N. Chancellor and H. Kusumaatmaja (2025), Integrating quantum algorithms into classical frameworks: a predictor–corrector approach using HHL, \textit{Quantum Science Technology 10}
%\bibitem{Hara1992}
%T. Hara and Gordon Slade (1992), Self-Avoiding Walk in Five or More Dimensions, \textit{Communications in Mathematical Physics. 147, 101-136 (1992) Communications}.

\end{thebibliography}

\end{document}